\documentclass{PoS}
\def\third{{3$^{\rm rd}$}}
\def\deg{{$^{\circ}$}}
\def\bd17{\mbox{BD +17\deg 3248}}
\def\cs22{\mbox{CS 22892-052}}
\def\rpro{{\em r}-process}
\def\gtaprx{ \mathrel{  \vcenter{
                        \offinterlineskip \hbox{$>$}
                        \kern 0.3ex \hbox{$\sim$}    } } }

\title{r-Process Enhanced Halo Stars}

\ShortTitle{r-Process Enhanced Halo Stars}

\author{\speaker{John J. Cowan}\\
        Homer L. Dodge Dept. of Physics, University of Oklahoma
Norman, OK 73019\\
        E-mail: \email{cowan@nhn.ou.edu}}

\author{Christopher Sneden\\
        Department of Astronomy and
McDonald Observatory, University of Texas,
Austin, Texas 78712\\
        E-mail: \email{chris@verdi.as.utexas.edu}}

\author{James E. Lawler\\
        Department of Physics, University of Wisconsin, Madison,
Madison, WI 53706 \\
        E-mail: \email{jelawler@wisc.edu}}

\author{ Elizabeth A. Den Hartog\\
        Department of Physics, University of Wisconsin, Madison,
Madison, WI 53706\\
         E-mail: \email{eadenhar@wisc.edu}}

\abstract{
Abundance observations indicate the presence of 
rapid-neutron capture ({\it i.e.}, $r$-process) elements in 
old Galactic halo and globular cluster stars. These 
observations provide insight into the nature of the earliest 
generations of stars in the Galaxy -- the progenitors of the 
halo stars -- responsible for neutron-capture synthesis of 
the heavy elements. 
The large 
star-to-star scatter observed in the abundances of 
neutron-capture element/iron ratios at low metallicities -- 
which diminishes with increasing metallicity or [Fe/H] -- 
suggests the formation of these heavy elements (presumably 
from certain types of supernovae) was rare in the early 
Galaxy. The stellar abundances also indicate a change from 
the $r$-process to the slow neutron capture ({\it i.e.}, 
$s$-) process at higher metallicities in the Galaxy and 
provide insight into Galactic chemical evolution. Finally, 
the detection of thorium and uranium in halo and globular 
cluster stars offers an independent age-dating technique 
that can put lower limits on the age of the Galaxy, and 
hence the Universe. 
}

\FullConference{International Symposium on Nuclear Astrophysics --- Nuclei in the Cosmos --- IX\\
		 June 25-30 2006\\
		 CERN, Geneva, Switzerland}

\begin{document}

\section{Introduction}

Neutron-capture elements
are present in some of the oldest, most 
metal-poor ({\it i.e.}, low iron abundance) Galactic halo stars.
These elements are synthesized in either the slow ($s$) or rapid ($r$) 
neutron-capture process
in generations of stars preceding the halo stars\cite{sne03a,cow06}. 
The abundances of these $n$-capture elements, particularly 
where they are enhanced in 
the Galactic halo stars,  are being employed to 
provide clues
and constraints on  a number of topics.   
These topics include the nature of the synthesis and the identities of the
stellar generations - those that preceded the halo stars - 
in the early Galaxy. The stellar abundances are also providing new insight into
the astrophysical site (or sites) for $r$-process nucleosynthesis.
The $s$-process seems to 
be identified   
in asymptotic giant branch (AGB) stars \cite{bus99}. 
However, the astrophysical site for the $r$-process has not been determined, 
although it is likely in a supernova environment\cite{cow04}.
Neutron-capture 
abundance determinations in a wide variety of stars are  also helping 
in understanding the nature of, and the trends in, Galactic chemical 
evolution. 
The detections of the long-lived radioactive elements, such as 
thorium and uranium, provide  direct chronometric age
determinations  for the oldest stars, which  in turn place constraints 
on age estimates
for the Galaxy and the Universe.

\section{n-Capture Element Abundances in the Halo Stars}

Comprehensive, high resolution abundance studies have been made 
for a number of $r$-process enhanced halo stars, including
\cs22\cite{sne03b}, \bd17\cite{cow02}, HD 115444\cite{wes00}, 
CS 31082-001\cite{hill02} and HD 221170\cite{iva06}.

\subsection{Dependence Upon Atomic Data}

\begin{figure}[ht]
\centering
\includegraphics[angle=0,width=2.90in]{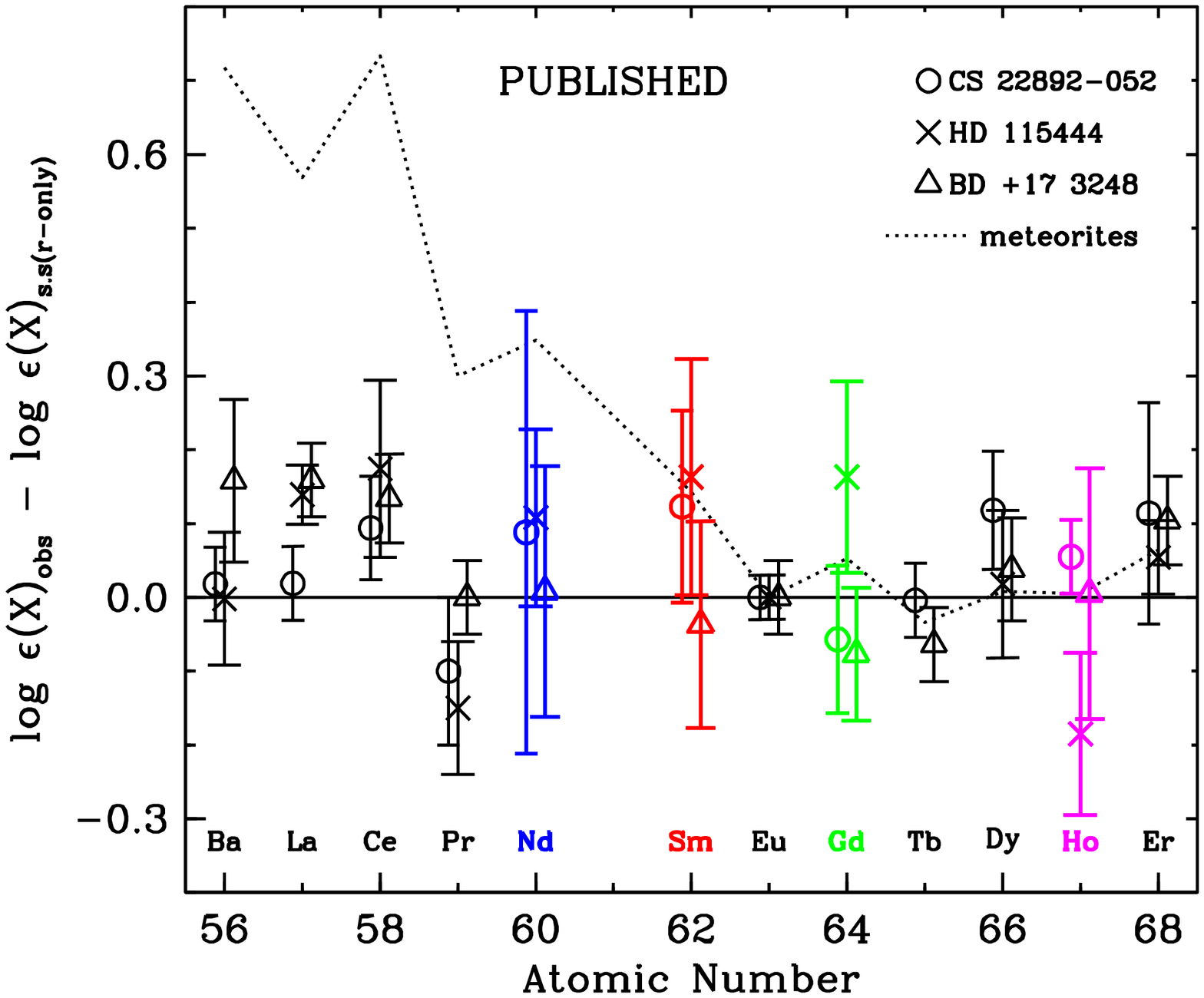}
\includegraphics[angle=0,width=2.90in]{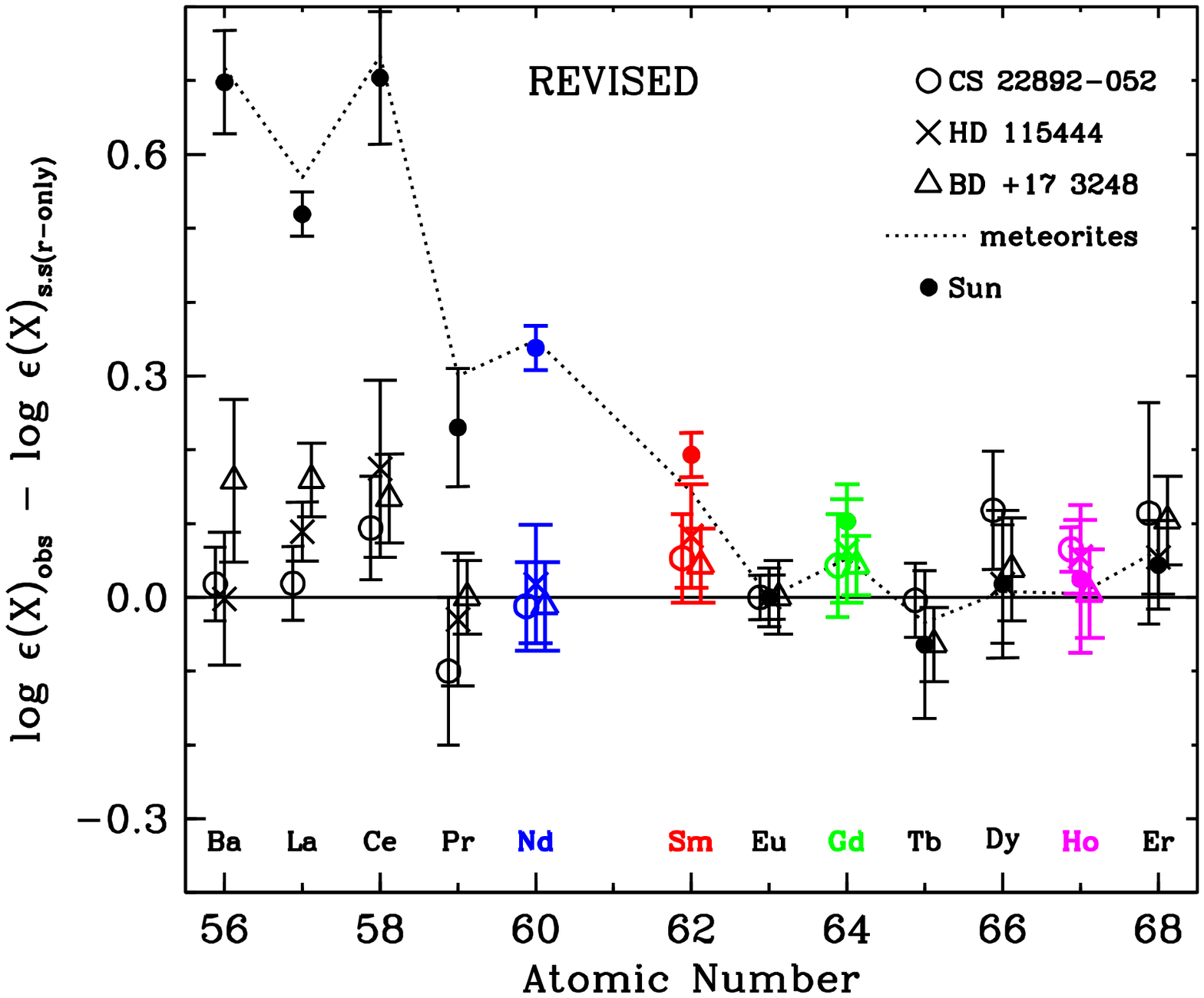}
\vskip -0.8in
\caption{(left) Abundance values (scaled at the element Eu)
for selected elements in the stars \cs22 (open circles), HD 115444 (x),
\bd17 (triangles) and the Sun (filled circles),
based upon previously published atomic
data.
(right) Newly derived abundance values based upon recent experimental lab data
(after \cite{law06,den06}).
\label{fig1}}
\end{figure}

The stellar abundance determinations depend critically upon 
the atomic data. 
We illustrate  
this  in 
Figure \ref{fig1}. 
In the left panel we show  
the abundances from Ba-Er,  based upon previously published atomic data, for
the metal-poor halo stars 
\cs22\cite{sne03b}, \bd17\cite{cow02}, HD 115444\cite{wes00}
and the Sun\cite{lod03}. 
The abundances are scaled with respect 
to the $r$-process element Eu. The solid horizontal line is given 
by  
log$\epsilon(X)_{observed}$- log$\epsilon(X)_{S.S. (r-only)} = 0$. 
The $r$-process-only values are determined
by the deconvolution of the solar abundances into individual 
$s$- and $r$-process contributions\cite{sim04,cow06b}. 
We focus
on the abundance determinations for Nd, Sm, Gd  and Ho in these
four stars.
It is clear in the (left panel of the) 
figure that there is a large amount of scatter for 
those  elements,  using the previously published atomic data. 
The right panel of Figure~\ref{fig1} 
shows the revised stellar and solar abundances employing new
atomic data.  
There has been a concerted effort in the last decade
to place these data on a firm, experimental basis. Newly measured 
values, particularly transition probabilities, 
have been obtained for the elements:
Nd\cite{den03},
Ho\cite{law04},
Pt\cite{den05},
Sm\cite{bie89,law06},
and Gd\cite{den06}.
It is clear from the figure that as a result of employing these newly
revised atomic data, the abundance  
scatter for the elements Nd, Ho, Sm and Gd has been dramatically reduced. 
This indicates that the
relative abundances of these elements are the same in all of these
metal-poor halo stars. 
We  note further that the abundances of {\it all} of the $n$-capture elements
in the  three stars are 
consistent with the SS $r$-process-only abundances -- the stellar abundance
values fall close to the solid horizontal line 
log$\epsilon(X)_{observed}$- log$\epsilon(X)_{S.S. (r-only)}$ = 0.

While the agreement is excellent, 
there are still some small deviations 
between the stellar and solar $r$-process abundances. 
The solar $r$-process values have been obtained as residuals, after 
first subtracting the $s$-process values from the total Solar System abundances.
Since the stellar abundances have become increasingly more reliable,
in large part due to the more accurate laboratory atomic data, 
it may be possible to predict directly 
the $r$-process abundances,
rather than relying on the subtraction of the $s$-process components
from the Solar System abundances.
This has been attempted recently for the elements 
Gd, Sm, Nd and Ho\cite{den06}. While there are still some 
uncertainties in the individual values, 
the 
technique could  be a viable new method of predicting the 
$r$-process-only contributions to
the Solar System abundances. 

\subsection{Observational Evidence for Two r-Processes in Nature?}

\begin{figure}[ht]
\centering
\includegraphics[angle=0,width=3.90in]{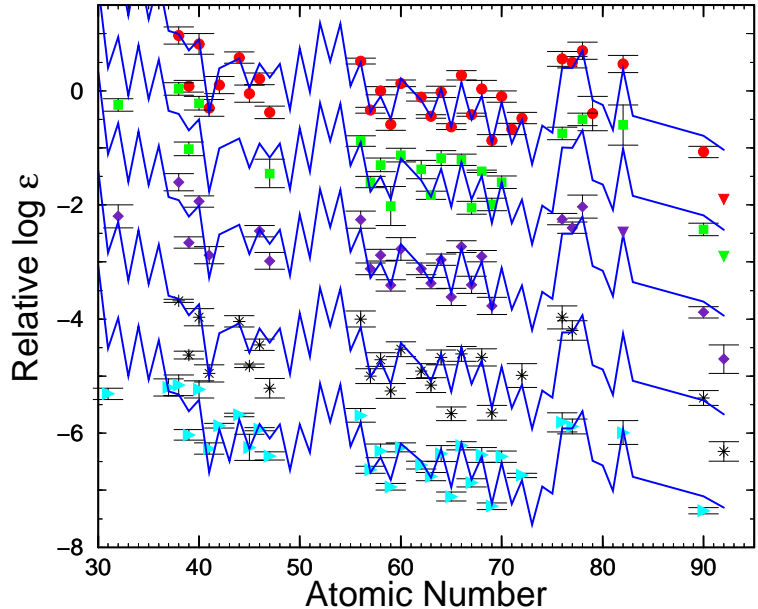}
\vskip -0.2in
\caption{An observational summary of the observed n-capture elements in 
five r-process rich halo stars: 
\cs22\ (filled circles, \cite{sne03b}),
HD~115444 (filled squares,\cite{wes00}),
\bd17\ (filled diamonds,\cite{cow02}),
CS~31082-001 (stars,\cite{hill02}),
and HD~221170 (filled
triangles,\cite{iva06}).
The vertical scale for the \cs22\ abundance set is true, and abundances
of all of the other stars have been vertically scaled downward for
display purposes.
Each of these stellar abundance sets is overlaid with the scaled
solar system \rpro\ abundance distribution that best fits the
observed abundances (solid lines).
\label{fig2}}
\end{figure}

Stellar abundances of the enhanced, metal-poor Galactic halo stars have
shown a consistently solar $r$-process pattern for the stable elements
Ba and above\cite{sne03a,cow06}. 
It has only been in recent years, however, that elemental abundances
between Sr-Y-Zr and Ba have been detected in these stars\cite{sne03b}. 
We illustrate in 
Figure~\ref{fig2} a summary of the observational data for the five 
$r$-process enhanced stars 
\cs22\cite{sne03b} , \bd17\cite{cow02}, HD 115444\cite{wes00}, 
CS 31082-001\cite{hill02} and HD 221170\cite{iva06}. 
The data have been scaled vertically (except for \cs22) for illustration
purposes. A scaled solar system $r$-process only abundance curve 
(solid line) is 
superimposed on each set of stellar data.
While some of these stars (e.g., HD 115444) have little data between 
Z=40 and Z=50, most of these lighter $n$-capture element abundances seem 
to lie somewhat below the solar system curves that fit the 
heavier $n$-capture data.
In particular Ag is below the SS curve in all cases. 
This seems to suggest a different origin or perhaps a different site for 
the synthesis of these two groups of $n$-capture elements, and in particular
may support the suggestion (based upon solar system meteoritic data) 
that there are two $r$-process sites in nature\cite{was96}. 
Whether there are actually two separate supernova sites\cite{was00} or
just different regions of the same supernova site\cite{cam01}
is still unclear. If there are two sites for this synthesis, it is also 
unclear at what element the split occurs\cite{kra06}. 
Clearly more observational and theoretical studies will be required to 
better understand the origin of both the lighter and heavier 
$n$-capture elements. 

\section{Chemical Evolution of the Elements}

\begin{figure}[ht]
\centering
\includegraphics[angle=0,width=2.90in]{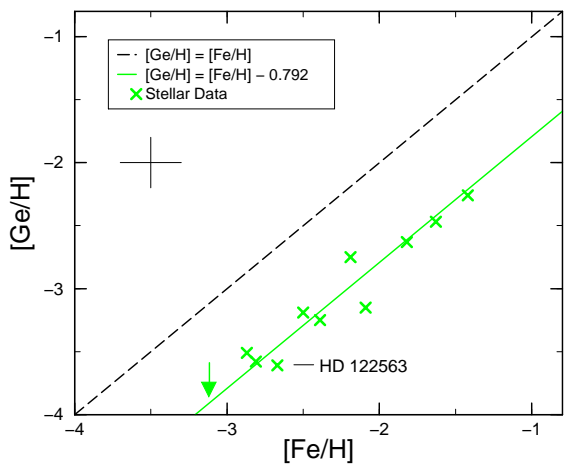}
\vskip -0.2in
\caption{
Relative abundances of the elemental ratio [Ge/H] as a function of
metallicity ({\it i.e.}, Fe) for a sample of 11 Galactic halo stars 
(after \cite{cow05}).
The arrow represents the derived upper limit for \cs22.
The dashed  line indicates the solar abundance ratio of these elements,
[Ge/H]~=~[Fe/H], while the solid green line shows the derived correlation
[Ge/H]~= [Fe/H]~--~0.79.
A typical error is indicated by the cross.
\label{fig3}}
\end{figure}

In the last few years observations using the STIS of the HST have opened up 
the element range available for study in these halo stars. In particular
elements such as Ge have their strongest transitions in the UV part of 
the spectral region, 
and thus are not available to ground-based observations. 
A recent survey of this element (along with Zr, Os, Ir and Pt) in a group of 11 
metal-poor, $n$-capture enhanced Galactic halo stars was undertaken\cite{cow05}.
The results are shown for Ge in Figure~\ref{fig3}.  
Surprisingly,  it appears that the Ge abundances in these stars correlate with
the iron abundance, with a ratio of [Ge/H] $\simeq$  [Fe/H]  - 0.8.  
Since Ge is normally thought of as an $n$-capture element, this behavior seemed
initially surprising. It may be, however, that for these most metal-poor  
stars, some type of charged-particle reactions could  be responsible for the
earliest Galactic synthesis of Ge\cite{fro05}, 
before the eventual $s$-process
contribution sets in.

\begin{figure}[ht]
\centering
\includegraphics[angle=0,width=3.90in]{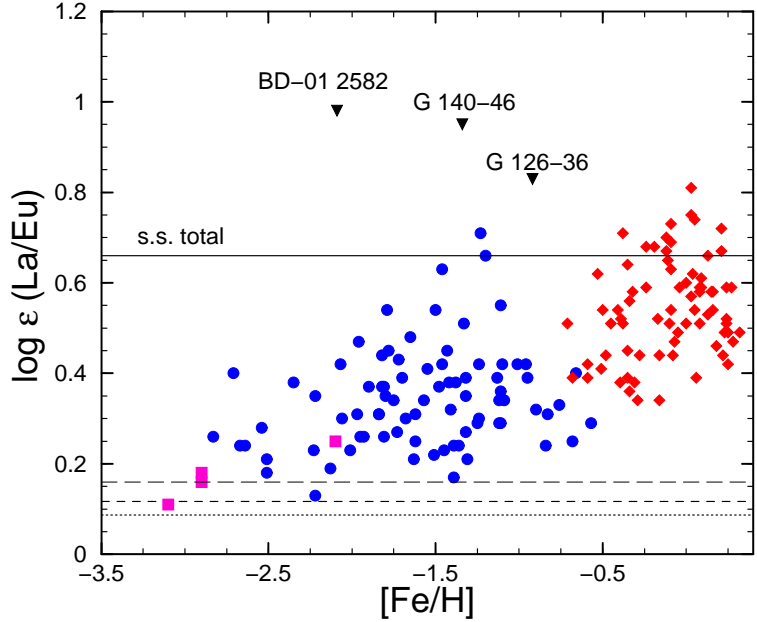}
\vskip -0.2in
\caption{
Abundance trends with respect to metallicity
for the elemental ratio La/Eu in a large number of halo 
(blue circles) and 
disk stars (red diamonds) in our
Galaxy (after \cite{sim04}). 
The long-dashed\cite{win06}, dashed\cite{obr03} and 
dotted\cite{bur00}  
lines are predictions for $r$-process only ratios by different 
investigators, and
the solid line is the total Solar System abundances.
The labeled stars are all $s$-process rich. The magenta squares are 
well-studied $r$-process enhanced stars. 
\label{fig4}}
\end{figure}

Certain chemical evolution trends can be understood with comparisons of 
$r$- and $s$-process elements as a function of metallicity - very roughly a 
time relationship. 
While the ($s$-process) element Ba has been used in ratio to the 
($r$-process) element Eu\cite{bur00},  
recent studies have employed La/Eu\cite{sim04}.
We illustrate in Figure~\ref{fig4}
the abundances  [La/Eu] in a group of Galactic halo\cite{sim04} (blue 
circles) and
disk\cite{wol95} (red diamonds) stars as a function of metallicity.  
Superimposed on the abundance data are predictions for the 
$r$-process only ratios: dotted line\cite{bur00}, dashed line\cite{obr03}
and long-dashed line\cite{win06}. 
(An additional value would fall between the dashed line and the
long-dashed line,  
based upon a very recent experimental determination of 
the 
neutron capture cross section of $^{139}$La, S. Marrone, private 
communication.) 
The magenta squares in Figure~\ref{fig4} are the $r$-process rich stars 
\cs22, HD 115444, CS 31082-001 and \bd17. 
There are several important points  worth 
noting in this  figure.
The general upward trend in the abundance ratio with increasing 
[Fe/H] reflects the increased contribution from the $s$-process to 
La production -- lower-mass stars have time to evolve and enrich the
interstellar gas with $s$-process-rich ejecta. 
It is also seen  
that at very low metallicities some of the ratios of La/Eu appear 
to lie above the $r$-process only value. This suggests some  $s$-process
contributions even at very low metallicities and, presumably, early
Galactic time.  
However, the onset of the bulk Galactic $s$-process nucleosynthesis 
occurs at somewhat
higher metallicity, but still below [Fe/H] = --2\cite{bur00,sim04}.  
Exactly where this main $s$-process production begins -- with implications
for the identities of the (mass ranges of) 
sites and nature of $s$-process synthesis -- 
depends upon the 
value of the $r$-process only ratio of La/Eu. This illustrates the  
critical nature and importance of the nuclear experimental data with
regard to the neutron cross-section measurement of $^{139}$La,
which in turn determines the $s$- and $r$-process 
components for this element. 

\begin{figure}[ht]
\centering
\includegraphics[angle=0,width=3.90in]{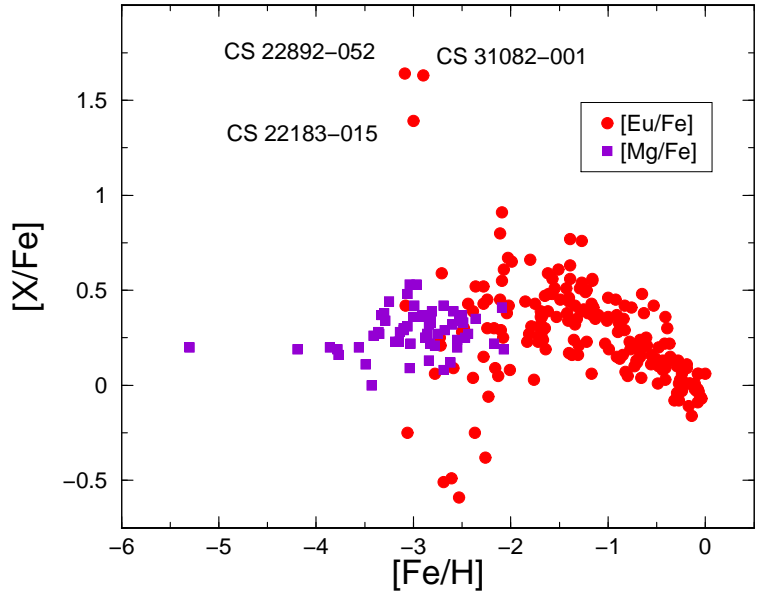}
\vskip -0.2in
\caption{
Abundance scatter of elemental ratios
[Eu/Fe] (\cite{sim04,wol95})
and [Mg/Fe] (\cite{cay04,arn05})
versus metallicity  for
samples of halo and
disk stars in our Galaxy 
(after \cite{cow06})
\label{fig5}}
\end{figure}

Additional information regarding the synthesis in the earliest 
generations of stars can be obtained by comparing 
elemental production as a function of metallicity.
We illustrate this behavior for Eu and Mg in 
a group of Galactic halo and disk stars
in 
Figure~\ref{fig5}\cite{cow06}.  
The large scatter in the Eu/Fe  abundance at very low metallicities
suggests that the Galaxy was chemically inhomogeneous ({\it i.e.}, unmixed) 
in $n$-capture elements at early times. 
At later times, and higher metallicities, this scatter diminishes. 
In contrast the ($\alpha$-element) Mg abundance data\cite{cay04,arn05} 
does not show this scatter. 
These abundance comparisons suggest 
that $r$-process synthesis (whatever the site) was a relatively rare event,
while Mg  production must have been more common at early Galactic times. 
These comparisons further suggest
possibly different sites for the major Eu and Mg synthesis. 
Since Mg is commonly thought to be produced in massive stars, Eu might
instead be synthesized in less massive stars that become 
supernovae\cite{cow04}.  

\section{Chronometers}

The detection of Th, with its very long half-life, 
allows for chronometric 
estimates for the ages of the halo stars.
The preferred  chronometer pair for these
age determinations is
U/Th, since both elements are made entirely in the $r$-process and they 
are nearby in  mass number (see discussions in \cite{sne03a,cow04,cow06}).
U/Th  has been employed  to determine the ages 
of CS 31082-001 (12.5 $\pm$ 3 Gyr,\cite{cay01}; 
14 $\pm$ 2.4 Gyr\cite{hill02}, ; 15.5 $\pm$ 3.2 Gyr\cite{sch02}, 
 14.1 $\pm$ 2.5 Gyr\cite{wan02},) and 
\bd17 (13.8 $\pm$ 4 Gyr\cite{cow02}).  
However,   uranium (with its low abundance and frequent 
spectral blending with molecules)
is difficult to detect in most stars.
 
Despite being widely separated in mass number from Th, 
Eu is 
easily detected  with  ground-based observations and made 
almost entirely in the $r$-process\cite{sim04}.
For these reasons 
chronometric age estimates  based upon the Th/Eu ratio 
have been made for a number of halo stars, 
typically
with age ranges of 11--15 Gyr
\cite{sne96,pfe97,cow99,sne00a,wes00,joh01,cow02,sne03b,iva06}
and with average age uncertainties $\simeq$ 3-4 Gyr.
This approach has also yielded a similar age for one globular
cluster\cite{sne00b}.
These stellar values are also consistent with age determinations for the Galaxy
based upon main-sequence turn-off ages for globular
clusters\cite{kra03} and
recent cosmological age estimates from WMAP\cite{teg04}
and Type Ia SNe\cite{rie98,per99}.

We note, however, that the 
chronometric age estimates depend sensitively upon 
the initial predicted values of Th/Eu, which in turn depend
on the nuclear mass formulae and $r$-process models
employed in making those
determinations (see discussion in \cite{sch02,cow04a}).
We also note that in at least one case, the chronometer ratio Th/Eu
does not appear to be usable.
While the abundances of the  stable
elements (through the \third $r$-process peak) in
CS 31082-01  
are consistent
with the scaled solar system $r$-process distribution, Th and U are
enhanced  
with respect to the other $n$-capture elements\cite{hill02,sch02}.
Thus, in this star the Th/Eu chronometer leads to  
unreasonably low age determinations, while the U/Th ratio 
predicts an age consistent
with other ultra metal-poor halo stars. 
However   
the lead abundance\cite{ple04}, which predominantly results from 
$\alpha$ decay of Th and U isotopes, 
in CS 31082-01  seems to be too low with respect
to such high abundance values of these radioactive elements\cite{kra04}.
Clearly, additional observational 
and theoretical studies of this
star,  and others with enhanced Th and U abundances,  will be needed to 
better  understand and further refine the chronometric age determinations.
                                                                                
\section{Conclusions}

Abundance studies of the old, metal-poor Galactic halo stars are providing
a wealth of data and information regarding the nature of 
early nucleosynthesis and the identities of the earliest stellar
generations. New clues to the site of the $r$-process are also coming
from such studies. 
Comparisons of the abundance levels in these 
stars,  with respect to younger and more metal-rich Galactic stars,  
provides insight into the history of chemical evolution in the Galaxy.
In particular the influence, and period of onset, of $s$-process nucleosynthesis 
in the Galaxy  are seen in these comparisons.
The detection   of 
the radioactive elements (such as Th and U) 
provides a technique to determine the ages of the
oldest stars,  which will help to determine the age of the Galaxy and
the Universe.

\acknowledgments
We thank our colleagues for valuable insights and contributions.
This work has been supported in part by NSF
grants AST 03-07279 (J.J.C.),
AST 03-07495 (C.S.),
AST 05-06324 (J.E.L.),
and by STScI. JJC thanks the 
Institute for Nuclear
Theory at the University of Washington
for its hospitality and the Department of Energy for
partial support during the
period when this paper was initiated.

\end{document}